\documentclass[%
 reprint,
groupedaddress,
showpacs,
 amsmath,amssymb,
 aps,
prl,
]{revtex4-1}

\usepackage{graphicx}
\usepackage{dcolumn}
\usepackage{bm}

\begin{document}

\newcommand{\be}{\begin{equation}}
\newcommand{\ee}[1]{\label{#1}\end{equation}}
\newcommand{\bem}{\begin{eqnarray}}
\newcommand{\eem}[1]{\label{#1}\end{eqnarray}}
\newcommand{\eq}[1]{Eq.~(\ref{#1})}
\newcommand{\Eq}[1]{Equation~(\ref{#1})}
\newcommand{\vp}[2]{[\mathbf{#1} \times \mathbf{#2}]}


\title{Comment on ``Supercurrent in a room temperature Bose-Einstein magnon condensate''}


\author{E.  B. Sonin}
 \affiliation{Racah Institute of Physics, Hebrew University of
Jerusalem, Givat Ram, Jerusalem 91904, Israel}

\date{\today}

\begin{abstract}
The comment explains that the preprint arXiv:1503.0042 has not presented persuasive theoretical or experimental arguments of existence of spin supercurrents 
in a magnon condensate prepared in a room temperature yttrium-iron-garnet magnetic film because the authors did not check known criteria for existence of spin supercurrents in magnetically ordered materials. Also they did not compare their supercurrent interpretation with a competing and more realistic scenario of  transport by spin diffusion.
        \end{abstract}

 \maketitle

The authors of the  preprint \cite{spinY} declared experimental evidence for the existence of a spin supercurrent in a coherent magnon condensate in a room temperature yttrium-iron-garnet magnetic film. Spin supercurrent is a manifestation of spin superfluidity, which has already been discussed from 70s of the last century (see Ref.~\onlinecite{Adv} and references to older papers therein). The theoretical analysis of this interesting phenomenon continues nowadays \cite{Halp,Tserk,Loss}. Experimental observation of it in magnetically ordered solids would be an essential breakthrough in the condensed matter physics. But I argue in this Comment that unfortunately  the claim on detection of spin supercurrent is premature and has not been sufficiently supported by presented experimental results and their theoretical interpretation.

The main difference of spin supercurrent from a more common spin diffusion current is that the latter is proportional to the gradient of spin magnetization, while the former is proportional to the  gradient of  the phase $\varphi$ (spin rotation angle in a plane) and is not accompanied by dissipation. However, not any spin current proportional to  $\nabla\varphi$ is a manifestation of spin superfluidity and deserves the title ``supercurrent''. The current $\propto\nabla\varphi$ exists in any spin wave and in any domain wall and can be generated by some disorder. In all this cases variation of the phase $\varphi$ is small (very small in weak spin waves and not more then on the order $\pi$ in domain walls and in disordered materials). Analogy with mass and charge persistent currents (supercurrents) emerges when  
at long (macroscopical) spatial intervals along streamlines the phase variation is many times larger than $2\pi$. Only this justifies calling it {\em macroscopic quantum phenomenon} as mass superfluidity is usually named. 

One might consider it as a purely semantic issue. Indeed, there is no law in the books, which forbids to call any current $\propto\nabla\varphi$  as a supercurrent. 
However, accepting such a nomenclature would reduce spin supercurrent to a trivial ubiquitous object. It does not worth efforts to perform new experiments on its detection, because decades-long observations of spin waves or domain structures confirm ``supercurrent'' existence beyond reasonable doubt.  Later on I keep in mind only a non-trivial 
spin supercurrent.

According to papers cited above, there are serious restrictions on existence of steady spin supercurrent: 

(i) The  gradient energy $\propto \nabla\varphi^2$ must exceed anisotropy energy in the easy plane, which prevents free rotation of phase in the plane.  This anisotropy is related to terms in the hamiltonian, which depend on the phase itself but not on its gradient.  

(ii) The gradient energy must not exceed the energy of  axial anisotropy, which keeps spin (or another vector in the spin space described by the angle $\varphi$) in the easy plane. This restriction is an analog of the Landau criterion for mass supercurrents.

Thus spin superfluidity is possible only in easy-plane magnetically ordered systems with a sufficient window determined by these two criteria. This means that coherent magnon condensate, also called magnon BEC, does not automatically   lead to spin supercurrent existence. Yttrium-iron-garnet magnetic films have no evident easy plane
and the aforementioned window of spin supercurrent  existence is absent, or at least rather narrow. Therefore it was concluded that observation of  spin supercurrents  in magnon condensates in this material is hardly possible (see the end of Sec.~2.9 in Ref.~\onlinecite{Adv}). In the light of it the report on observation of spin supercurrent in yttrium-iron-garnet magnetic films arose at least questions. The preprint did not check the criteria of spin supercurrent existence and did not even mention them or anisotropies connected with these criteria.

Another complication is that the authors investigated not stationary but time dependent spin currents. It is well known that in AC experiments on superconductivity or  superfluidity  supercurrents are always accompanied by dissipative normal currents. In magnon condensates an analog of normal current is spin diffusion current. It is proportional to the same  gradient of spin precession frequency, which determines linear growth of the phase in the preprint.  A stationary supercurrent is impossible in the presence of spin precession frequency gradient (analog of electric field in superconducting metals or of chemical potential gradient in neutral superfluids) because these gradients accelerate supercurrents.
The main (and only) argument of the authors in favor of their claim is ``surprisingly  high quantitative agreement between measured data and the analytical model''. The agreement looks very good  indeed, probably due to using a fitting parameter. But this is based on the assumption that in a magnon condensate there is no mechanism of spin transport other than supercurrent.  The authors did not consider a more common scenario of spin diffusion, which could probably successfully  compete with their present interpretation. It is worth noting that spin diffusion would be more effective in the condensed state than in a non-condensed gas of magnons.

In summary, the preprint has not provided persuasive  evidence of spin supercurrent in yttrium-iron-garnet magnetic films because: (i) they did not check known criteria for existence of spin supercurrent in this material; (ii) they did not discuss or estimate a more natural scenario that spin transport in their experiment was purely diffusive.

I acknowledge very useful discussions with Victor L'vov.


%

\end{document}